\documentclass[sigconf]{acmart}
\usepackage{multirow}
\usepackage{colortbl}
\usepackage{listings}
\usepackage{tcolorbox}
\tcbuselibrary{listings,breakable,skins}
\renewcommand{\footnotetextcopyrightpermission}[1]{}
\AtBeginDocument{%
  }

\setcopyright{acmlicensed}
\copyrightyear{2018}
\acmYear{2018}
\acmDOI{XXXXXXX.XXXXXXX}
\acmConference[Conference acronym 'XX]{Make sure to enter the correct
  conference title from your rights confirmation email}{June 03--05,
  2018}{Woodstock, NY}
\acmISBN{978-1-4503-XXXX-X/2018/06}

\begin{document}

\title{SHARP: Semantic Hierarchical Agentic Reasoning for Next POI Prediction with Dynamic Memory}

\author{Yuhang Zhan}
\affiliation{%
  \institution{University of Electronic Science and Technology of China}
  \city{Chengdu}
  \country{China}
}
\email{agooostoo@gmail.com}

\author{Silin Zhou}
\affiliation{%
  \institution{University of Electronic Science and Technology of China}
  \city{Chengdu}
  \country{China}
}
\email{zhousilinxy@gmail.com}

\author{Zhuo Chen}
\affiliation{%
  \institution{University of Electronic Science and Technology of China}
  \city{Chengdu}
  \country{China}
}
\email{suansuannnn@gmail.com}

\author{Haoxiang Lei}
\affiliation{%
  \institution{University of Electronic Science and Technology of China}
  \city{Chengdu}
  \country{China}
}
\email{haoxianglei26@gmail.com}

\author{Yunren Feng}
\affiliation{%
  \institution{University of Electronic Science and Technology of China}
  \city{Chengdu}
  \country{China}
}
\email{yunrengfeng@gmail.com}

\author{Lisi Chen}
\affiliation{%
  \institution{University of Electronic Science and Technology of China}
  \city{Chengdu}
  \country{China}
}
\email{lchen012@e.ntu.edu.sg}

\author{Shuo Shang}
\affiliation{%
  \institution{University of Electronic Science and Technology of China}
  \city{Chengdu}
  \country{China}
}
\email{jedi.shang@gmail.com}

\renewcommand{\shortauthors}{Zhan et al.}

\begin{abstract}
Next point-of-interest (POI) prediction supports personalized location services by
anticipating users' next destinations. Recent methods leverage large
language models (LLMs) for semantic reasoning and graphs for collective
mobility modeling, improving POI prediction. However, three challenges
still remain. Bridging LLMs' semantic reasoning and specific POI prediction remains challenging.
User mobility also evolves
over time, making it necessary to retain and update historical knowledge
within limited LLM contexts. Meanwhile, global transition graphs smooth
out time-specific category transition patterns.
To address these challenges, we propose \textbf{S}emantic
\textbf{H}ierarchical \textbf{A}gentic \textbf{R}easoning for Next POI
\textbf{P}rediction with Dynamic Memory (SHARP), an agentic hierarchical
framework in which category-level reasoning guides specific POI
prediction. A trajectory memory agent maintains a dynamic trajectory
memory bank that preserves earlier mobility patterns and tracks
behavioral changes with limited context growth. We construct
multi-granularity time-aware category transition graphs with global
and time-conditioned views to preserve temporal regularities obscured
by global aggregation. A category predict agent combines these graphs
with trajectory memory to predict likely next categories.
SHARP learns semantic IDs (SIDs) through a residual-quantized
variational autoencoder (RQ-VAE) to construct semantic POI
representations, which are jointly optimized with a category-guided
POI predictor end-to-end.
On three real-world datasets, SHARP improves Acc@5 and Acc@10 by an
average of 5.29 and 4.52 percentage points, respectively, over the
strongest baseline.
\end{abstract}

% \keywords{Next POI prediction, LLM agents, Dynamic trajectory memory,
%   Time-aware category transition graphs}

\maketitle

\section{Introduction}

\begin{figure}[t]
  \centering
  \newsavebox{\challengeleft}
  \newsavebox{\challengecaption}
  \begin{lrbox}{\challengeleft}
  \begin{minipage}[b]{0.64\columnwidth}
    \centering
    \includegraphics[width=\linewidth]{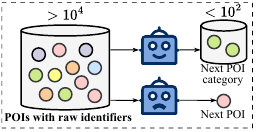}\par
    {\small\bfseries (a) Challenge 1: Connecting Semantic Reasoning
    with Specific POI Prediction.\par}
    \medskip
    \includegraphics[width=\linewidth]{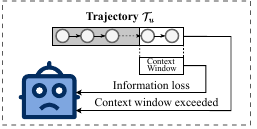}\par
    {\small\bfseries (b) Challenge 2: Capturing Evolving Mobility
    within Limited Contexts.\par}
  \end{minipage}
  \end{lrbox}
  \begin{lrbox}{\challengecaption}
  \begin{minipage}[b]{0.33\columnwidth}
    \centering
    {\small\bfseries (c) Challenge 3: Preserving Time-Conditioned
    Transition Patterns.\par}
  \end{minipage}
  \end{lrbox}
  \begin{minipage}[t]{0.64\columnwidth}
    \vspace{0pt}
    \usebox{\challengeleft}
  \end{minipage}\hfill
  \begin{minipage}[t][\dimexpr\ht\challengeleft+\dp\challengeleft\relax][s]{0.33\columnwidth}
    \vspace{0pt}
    \centering
    \includegraphics[
      width=\linewidth,
      height=\dimexpr\ht\challengeleft+\dp\challengeleft-\ht\challengecaption-\dp\challengecaption-\smallskipamount\relax,
      keepaspectratio
    ]{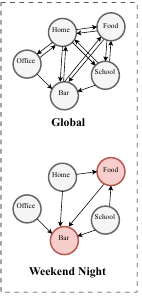}\par
    \vfill
    \nointerlineskip
    \usebox{\challengecaption}
  \end{minipage}
  \caption{Three challenges in next POI prediction.}
  \Description{The upper-left panel illustrates predicting a POI from
  raw identifiers. The lower-left panel contrasts historical information
  loss with exceeding the LLM context window. The right panel contrasts
  global and weekend-night transition patterns using schematic graphs.}
  \label{fig:intro-comparison}
\end{figure}

The widespread use of location-based social networks and mobile location
services has generated extensive data about points of interest (POIs)
and users' visits to them~\cite{yang2022getnext,feng2018deepmove}.
Platforms such as Foursquare provide user check-ins~\cite{yang2015foursquare},
while mobile applications collect location records~\cite{feng2018deepmove}.
POI attributes and user reviews further describe the places users
visit~\cite{gao2015contentaware}.
This wealth of data enables the study of individual preferences and
collective mobility, making next POI prediction an important task for
anticipating where people will go. Such predictions support applications
such as personalized location recommendation~\cite{gao2015contentaware},
intelligent transportation~\cite{feng2018deepmove}, and urban
planning~\cite{feng2018deepmove}.
Next POI prediction aims to predict the POI a user is most likely to visit next
given their past check-ins and available context, accounting for
individual preferences, sequential dependencies, and spatiotemporal
factors.

Recent methods advance next POI prediction through sequential modeling,
graph learning, and semantic reasoning. DeepMove~\cite{feng2018deepmove}
and STAN~\cite{luo2021stan} capture historical regularities and
spatiotemporal relations, while GETNext~\cite{yang2022getnext} and
STHGCN~\cite{yan2023sthgcn} model shared transitions through graphs,
enabling prediction to incorporate population-level mobility patterns.
Beyond these behavioral signals, LLM-Mob~\cite{wang2023llmmob} and
LLM4POI~\cite{li2024llm4poi} bring the prior knowledge and semantic
reasoning of large language models (LLMs) into prediction.
AgentMove~\cite{feng2025agentmove} and CoMaPOI~\cite{zhong2025comapoi}
use agentic reasoning and collaboration to integrate complementary
mobility information, while
GNPR-SID~\cite{wang2025gnprsid} and SPAR~\cite{wang2026spar} encode
POI relations through semantic IDs for generative prediction,
allowing related POIs to share semantic information through their identifiers.
Despite these advances, three challenges remain.
(1) Connecting LLMs’ semantic reasoning with specific POI prediction remains challenging.
(2) A user's mobility patterns evolve over time, requiring earlier routines
and preferences to be considered alongside recent behavioral changes
within limited LLM contexts.
(3) Global POI transition graphs smooth out local regularities under
different temporal conditions, obscuring patterns relevant to prediction.

\noindent{\textnormal{\bfseries Challenge 1: Connecting Semantic Reasoning with Specific POI Prediction.}}
Directly predicting specific POIs with LLMs~\cite{wang2023llmmob,li2024llm4poi,zhong2025comapoi}
places the entire prediction task on the LLM, requiring it to turn
semantic reasoning about mobility into a specific POI prediction. This
mapping is difficult, especially when the outputs are raw identifiers
whose meanings are not accessible from the identifiers themselves.
Figure~\ref{fig:intro-comparison}(a) contrasts this direct prediction
with predicting a POI category. Category names such as Restaurant and
Shopping Mall have recognizable meanings that align with LLMs' prior
knowledge and semantic reasoning, making category prediction a
substantially easier task for LLMs than directly predicting specific POIs.
This motivates separating category prediction from specific POI
prediction within a hierarchical framework, where LLMs predict
categories to guide subsequent POI prediction. Semantic POI
representations connect the LLM's semantic guidance with the POI
predictor, rather than requiring the LLM to produce the final POI itself.

\noindent{\textnormal{\bfseries Challenge 2: Capturing Evolving Mobility within Limited Contexts.}}
A user's mobility patterns evolve over time, but earlier trajectories
remain informative about recurring routines and
preferences~\cite{feng2018deepmove,zhong2025comapoi}. These
long-term patterns may be absent from recent check-ins, so considering
only recent trajectories provides an incomplete view of the user's
mobility. Retaining the complete history in LLM inputs, however, makes
input token consumption grow linearly with accumulated check-ins and
eventually exceed the context window~\cite{feng2025agentmove,zhong2025comapoi}.
Figure~\ref{fig:intro-comparison}(b)
illustrates this tension between retaining historical information and
keeping the input within the available context. This motivates a dynamic
trajectory memory bank that preserves earlier mobility patterns and
incorporates changes revealed by new trajectories. By carrying this
memory forward, the agents can interpret recent behavior in light of
long-term patterns without repeatedly processing the full history,
limiting context growth and token consumption.

\noindent{\textnormal{\bfseries Challenge 3: Preserving Time-Conditioned Transition Patterns.}}
Graph-based methods such as GETNext and STHGCN capture collective
mobility by graph representations~\cite{yang2022getnext,yan2023sthgcn}.
POI-GDE further incorporates time through time-serial graphs and
interval-aware updates~\cite{yang2024poigde}. Global aggregation,
however, can smooth out regularities prominent under specific time
conditions, as illustrated by the weekend-night transitions in
Figure~\ref{fig:intro-comparison}(c). Even when temporal information is
encoded, latent graph embeddings do not explicitly convey transition
semantics that LLMs can directly interpret. Preserving temporal
regularities and making them accessible to LLM reasoning therefore
requires explicit semantic descriptions of these transitions. We
construct multi-granularity time-aware category transition graphs,
preserving global and time-conditioned patterns in a semantically
interpretable form for LLM reasoning.

Building on these insights, we propose \textbf{S}emantic
\textbf{H}ierarchical \textbf{A}gentic \textbf{R}easoning for Next POI
\textbf{P}rediction with Dynamic Memory (SHARP), an agentic hierarchical
framework that connects category-level reasoning with semantic POI
prediction. A trajectory memory agent maintains a dynamic trajectory
memory bank that preserves long-term mobility patterns and captures
recent behavioral changes without repeatedly processing the complete
history. Multi-granularity time-aware category transition graphs
complement this memory with global and time-conditioned population-level
patterns in a form accessible to LLM reasoning. 
Unlike CoMaPOI's direct POI prediction~\cite{zhong2025comapoi}, our category predict agent combines memory and graph information to predict categories that guide POI scoring without restricting eligibility.
For specific POI prediction, we learn semantic IDs (SIDs) through a residual-quantized variational autoencoder (RQ-VAE)~\cite{lee2022rqvae}. Unlike GNPR-SID's autoregressive SID generation~\cite{wang2025gnprsid}, we use the selected codebook vectors to construct semantic POI representations.
A category-guided POI predictor integrates these
representations with agent predictions, trajectory information, graph
information, and individual mobility patterns to predict the next
POI. The semantic POI representations and the POI predictor are
optimized jointly end-to-end.

Experiments on NYC, TKY, and CA show that SHARP outperforms twelve
baselines, improving Acc@5 and Acc@10 over the strongest baseline by
an average of 5.29 and 4.52 percentage points, respectively. Dynamic
trajectory memory reduces input token consumption by
86.14\%--91.86\% compared with Full History in our token analysis,
while improving category Acc@5 by 10.11--12.70 percentage points
over analyzing only the current trajectory.

Our contributions of this work are summarized as follows:
\begingroup
\settowidth{\leftmargini}{\textbullet}
\addtolength{\leftmargini}{\labelsep}
\begin{itemize}
  \item We propose SHARP, an agentic hierarchical framework that
  connects category-level reasoning with semantic POI representations.
  Jointly learning these representations and a category-guided POI
  predictor enables fine-grained prediction by agent reasoning.

  \item We develop a dynamic trajectory memory bank that preserves
  earlier mobility knowledge and tracks behavioral changes, supporting
  category prediction with substantially reduced token consumption.

  \item We construct multi-granularity time-aware category transition
  graphs that capture global and time-conditioned mobility patterns,
  providing temporally differentiated collective information for
  agentic category prediction.
\end{itemize}
\endgroup

\section{Related Work}

\noindent{\textnormal{\bfseries Non-LLM-Based Methods.}}
Methods without LLMs primarily model mobility through sequential
dependencies and collaborative relations.
PRME~\cite{feng2015prme} learns
personalized transition metrics, while recurrent
models~\cite{liu2016strnn,feng2018deepmove} capture spatiotemporal
dependencies. LSTPM~\cite{sun2020lstpm} models long-/short-term preferences,
while STAN~\cite{luo2021stan} and MTNet~\cite{huang2024mtnet}
capture spatiotemporal relations.
GETNext~\cite{yang2022getnext} combines graph information with a
Transformer~\cite{vaswani2017attention},
STHGCN~\cite{yan2023sthgcn} models higher-order relations,
and LoTNext~\cite{xu2024lotnext} addresses long-tailed visitation.
These approaches learn representations from behavioral and item
information, but do not exploit the broader semantic knowledge and
reasoning capabilities of pretrained LLMs.

\noindent{\textnormal{\bfseries LLM-Based Methods.}}
LLM-based methods use pretrained knowledge to reason about mobility.
LLM-Mob~\cite{wang2023llmmob} uses prompting,
while LLM4POI~\cite{li2024llm4poi} and
Refine-POI~\cite{li2025refinepoi} employ supervised
and reinforcement fine-tuning of LLMs, respectively.
AgentMove~\cite{feng2025agentmove} and CoMaPOI~\cite{zhong2025comapoi}
extend this direction through agentic reasoning and collaboration.
Semantic indexing also supports LLM-based generation.
LC-Rec~\cite{zheng2024lcrec} aligns item identifiers with language
and collaborative semantics, and GNPR-SID~\cite{wang2025gnprsid}
uses semantic IDs for generative POI prediction.
ROS~\cite{lv2026ros} and SPAR~\cite{wang2026spar} further incorporate
spatial knowledge into generative prediction.
Despite these advances, preserving evolving mobility patterns within
limited LLM contexts and integrating time-conditioned collective
transitions remain challenges.

% Declare the two-column overview early so it can appear with the method.
\begin{figure*}[t]
  \centering
  \includegraphics[width=\textwidth]{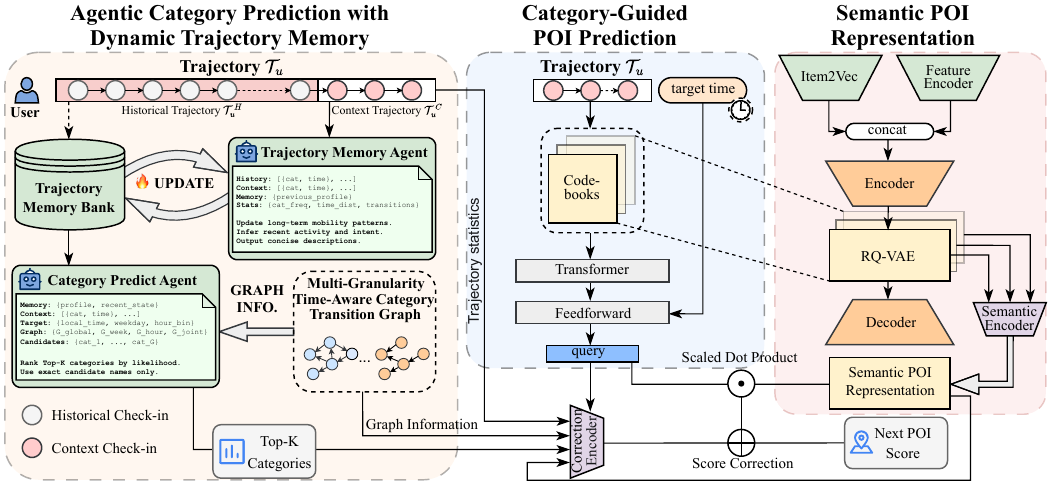}
  \caption{Overview of SHARP.}
  \label{fig:framework}
\end{figure*}

\section{Preliminary}

\noindent{\textnormal{\bfseries Next POI Prediction.}}
Let \(\mathcal{U}=\{u_1,\ldots,u_M\}\),
\(\mathcal{P}=\{p_1,\ldots,p_N\}\), and
\(\mathcal{G}=\{g_1,\ldots,g_Q\}\) denote the sets of users, POIs, and POI categories, respectively. Each POI
\(p_i\in\mathcal{P}\) is represented as a tuple
\(p_i=\left\langle \operatorname{id}_i,\operatorname{lon}_i,
\operatorname{lat}_i,g_i\right\rangle\),
where \(\operatorname{id}_i\) is its unique identifier,
\(\operatorname{lon}_i\) and \(\operatorname{lat}_i\) are its
longitude and latitude, and \(g_i\in\mathcal{G}\) is its category.

For a user \(u\in\mathcal{U}\), an observed check-in trajectory is a time-ordered sequence $\mathcal{T}_{u}
=\left\langle
(p_{u,1},t_{u,1}),(p_{u,2},t_{u,2}),\ldots,
(p_{u,n},t_{u,n})
\right\rangle$, where each pair $(p_{u,i},t_{u,i})$ records a check-in
by user $u$ at POI $p_{u,i}$ and timestamp $t_{u,i}$.
Given a split position \(1\leq m<n\), we divide $\mathcal{T}_{u}$ into
the historical trajectory $\mathcal{T}_{u}^{H}
=\left\langle
(p_{u,1},t_{u,1}),\ldots,(p_{u,m},t_{u,m})
\right\rangle$ and the context trajectory $\mathcal{T}_{u}^{C}
=\left\langle
(p_{u,m+1},t_{u,m+1}),\ldots,(p_{u,n},t_{u,n})
\right\rangle$, where $\mathcal{T}_{u}^{H}$ contains earlier check-ins
that characterize long-term mobility patterns, and $\mathcal{T}_{u}^{C}$
contains the most recent check-ins that reflect current mobility tendencies.
Given these two trajectories and a target time $t_{u,n+1}>t_{u,n}$,
the next POI prediction task aims to predict the POI
\(p_{u,n+1}\in\mathcal{P}\) that user $u$ will visit next.

\noindent{\textnormal{\bfseries Semantic IDs.}}
Conventional POI identifiers distinguish POIs but carry no information
about their relationships. A semantic ID (SID) instead represents an
item as a tuple of discrete codes learned from its representation, so
related POIs can share part of their SIDs~\cite{rajput2023tiger,wang2025gnprsid}. For example, two
similar POIs may be assigned
\(\langle 12,7,31\rangle\) and
\(\langle 12,7,18\rangle\), sharing the first two codes while
remaining distinguishable. Common SID construction methods include
Residual-Quantized Variational Autoencoder (RQ-VAE)~\cite{lee2022rqvae}, Residual
Quantization K-Means (RQ-KMeans)~\cite{wang2026spar}, and hierarchical
K-means~\cite{wang2022nci}. These
methods map continuous item representations into discrete code tuples.

\section{Method}
\label{sec:method}

\subsection{Framework Overview}

As illustrated in Figure~\ref{fig:framework}, SHARP follows a hierarchical
process that first predicts likely POI categories and then uses them to
guide specific POI prediction. Given historical and context trajectories,
the trajectory memory agent maintains a dynamic trajectory memory bank
that captures long-term mobility patterns and recent behavioral
tendencies. This memory is updated as new trajectories arrive. The
category predict agent combines the memory with the context trajectory,
target time, and information retrieved from multi-granularity time-aware
category transition graphs to predict likely next categories.

For specific POI prediction, SHARP uses semantic POI representations
constructed from SIDs learned through RQ-VAE. A category-guided POI
predictor encodes the trajectory using these representations and combines
it with the target time, agent category predictions, graph information,
and personalized mobility statistics to predict the next POI. This
connects category-level reasoning with distinctions among individual
POIs. The semantic POI representations and the prediction network are
optimized jointly end-to-end.

\subsection{Agentic Category Prediction with Dynamic Trajectory Memory}
\label{sec:agentic-category-prediction}

A user's mobility patterns evolve over time. Relying only on recent
trajectories overlooks recurring preferences and routines reflected in
earlier activities beyond those trajectories. Ignoring this longer-term
historical information leaves an
incomplete account of the user's mobility patterns and limits the
historical context available for interpreting current behavior. Effective user modeling
therefore requires considering longer-term mobility patterns alongside
recent behavior. Repeatedly feeding all
preceding trajectories to an LLM, however, makes the number of input tokens
per prediction grow linearly with the number of accumulated check-ins,
eventually exceeding the model's context window.
We therefore use a trajectory memory
agent to maintain dynamic trajectory memory and a category predict agent
to predict the next category using this memory, the target time, and
multi-granularity time-aware graph information.

\noindent{\textnormal{\bfseries Trajectory Memory Agent.}}
For user \(u\), the memory combines a long-term mobility profile
\(\mathbf{P}_{u,n}\), a context state \(\mathbf{C}_{u,n}\), and
personalized mobility statistics \(\mathbf{S}_{u,n}\):
\begin{equation}
\mathbf{M}_{u,n}
=\left\langle
\mathbf{P}_{u,n},\mathbf{C}_{u,n},\mathbf{S}_{u,n}
\right\rangle.
\label{eq:dynamic-user-memory}
\end{equation}
The long-term mobility profile summarizes recurring mobility routines, social
activity patterns, and category transitions, while the context state
describes recent activities and changes in these patterns. Statistical
tools~\cite{zhong2025comapoi} summarize the historical and context trajectories to obtain mobility
statistics:
\begin{equation}
\mathbf{S}_{u,n}
=\operatorname{Tool}_{\mathrm{stat}}\!\left(
\mathcal{T}_{u}^{H},\mathcal{T}_{u}^{C}
\right).
\end{equation}

As new historical trajectories become available, the trajectory memory
agent updates the previous long-term mobility profile using the current
historical trajectory and its statistics:
\begin{equation}
\mathbf{P}_{u,n}
=\operatorname{Agent}_{\mathrm{M}}^{\mathrm{long}}\!\left(
\mathcal{T}_{u}^{H},\mathbf{P}_{u,n-1},\mathbf{S}_{u,n}
\right),
\end{equation}
where \(\mathbf{P}_{u,n-1}\) is the previous profile, which is empty at
initialization. The agent
preserves recurring patterns and revises descriptions when new check-ins
indicate a change in behavior. This update carries earlier mobility
knowledge forward without requiring the
complete check-in history in every prompt.

The agent then interprets the context trajectory with reference to
the updated long-term mobility profile and mobility statistics:
\begin{equation}
\mathbf{C}_{u,n}
=\operatorname{Agent}_{\mathrm{M}}^{\mathrm{ctx}}\!\left(
\mathcal{T}_{u}^{C},\mathbf{P}_{u,n},\mathbf{S}_{u,n}
\right).
\end{equation}
Here \(\operatorname{Agent}_{\mathrm{M}}^{\mathrm{long}}\) and
\(\operatorname{Agent}_{\mathrm{M}}^{\mathrm{ctx}}\) denote the long-term mobility profile
update and context interpretation performed by the trajectory memory
agent. The context state captures current activity tendencies and how
they relate to the user's established routines. Updating the profile
and refreshing the context state allows the memory to track both
persistent behavior and recent changes.

\noindent{\textnormal{\bfseries Category Predict Agent.}}
The context trajectory provides the recent check-in sequence, while
\(\mathbf{M}_{u,n}\) provides the long-term mobility profile and context state.
The target time and the multi-granularity time-aware graph information
\(\mathbf{G}_{n}\) provide temporal and population-level transition patterns.
The graph construction and
retrieval are described in the next subsection.

The agent combines these inputs to predict the Top-\(K\) categories
from the category space \(\mathcal{G}\):
\begin{equation}
\begin{aligned}
\mathbf{g}_{u,n+1}^{K}
&=\left\langle
g_{u,n+1}^{(1)},\ldots,g_{u,n+1}^{(K)}
\right\rangle\\
&=\operatorname{Agent}_{\mathrm{Pred}}\!\left(
\mathcal{T}_{u}^{C},\mathbf{M}_{u,n},t_{u,n+1},\mathbf{G}_{n},
\mathcal{G}
\right),
\end{aligned}
\label{eq:agent-category-prediction}
\end{equation}
where \(g_{u,n+1}^{(k)}\in\mathcal{G}\). These predictions connect the
agents' interpretation of user mobility with subsequent POI prediction.
Prompt templates for both agents are provided in
Appendix~\ref{app:agent-prompts}.

\subsection{Multi-Granularity Time-Aware Category Transition Graph}
\label{sec:category-transition-graph}

A single macro-level transition graph aggregates all category
transitions into one population-level pattern. Although this graph
provides a stable estimate of common transitions, it ignores when a
transition occurs. The same source category may lead to different
activities on weekdays and weekends or at different times of
day~\cite{yang2015foursquare,huang2024mtnet}, and
aggregating these behaviors can obscure the transition pattern that is
most relevant to the current target time. We therefore model category
transitions through multiple temporal granularities rather than
relying on one global graph alone.

We divide time into two week types, weekday and weekend, and six
hour bins: late night \([0,6)\), morning \([6,11)\), noon \([11,14)\),
afternoon \([14,18)\), evening \([18,22)\), and night \([22,24)\).
These divisions induce 21 time-aware category transition graphs across
four views: one global graph, two week-type graphs, six hour-bin graphs,
and twelve joint graphs conditioned on both week type and hour bin.
The global view captures
stable transition patterns, while the conditional views describe specific temporal patterns.

All views share the category set \(\mathcal{G}\) as their nodes. Each
graph in these views is defined at time \(t\) as
\(G_t=\left(\mathcal{G},E_t,W_t\right)\),
where \(E_t\) denotes the directed edge set and
\(W_t\) contains the edge weights \(w_{ij}(t)\).
The graphs are constructed incrementally as check-ins arrive in
chronological order. Across all user trajectories, directed edges connect
the categories of consecutive check-ins, with weights representing
transition frequencies. These frequencies are accumulated in the global
graph and the conditional graphs selected by the destination check-in's
week type and hour bin.

\begin{figure}[t]
  \centering
  \includegraphics[width=\columnwidth]{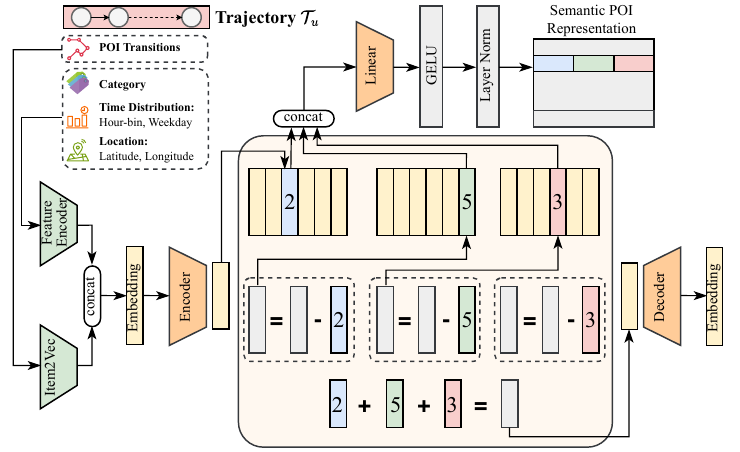}
  \caption{Semantic POI representation through RQ-VAE. Behavioral and
  structured POI representations are quantized into SIDs, and the selected
  codebook vectors are encoded into semantic POI representations.}
  \label{fig:sid-construction}
\end{figure}

A graph retrieval tool extracts one-hop and multi-hop category transition
distributions from the global graph and the time-conditioned graphs
selected according to the target time, using the final context category
as the source:
\begin{equation}
\mathbf{G}_{n}
=\operatorname{Tool}_{\mathrm{graph}}\!\left(
\mathcal{T}_{u}^{C},G_{t_{u,n}}
\right).
\end{equation}
These distributions provide multi-granularity transition patterns for
agentic category prediction and subsequent POI prediction.

\subsection{Semantic POI Representation}
\label{sec:semantic-poi-representation}

Raw POI identifiers distinguish individual POIs but do not express the
semantic relations between them. For POIs, these relations are reflected in
check-in sequences and their category, temporal, and geographic
characteristics. We therefore encode this information into a continuous
POI representation and discretize it into an SID through
RQ-VAE~\cite{lee2022rqvae,rajput2023tiger}, as illustrated in
Figure~\ref{fig:sid-construction}.
POIs with similar mobility characteristics can
therefore share part of their discrete representations.

We use Item2Vec~\cite{barkan2016item2vec} to learn behavioral POI representations from
co-occurrence patterns in chronologically ordered check-in sequences:
\begin{equation}
\mathbf{h}_p^{\mathrm{beh}}
=\operatorname{Item2Vec}(p).
\end{equation}

We then encode the concatenated structured POI features using a feature encoder:
\begin{equation}
\mathbf{h}_p^{\mathrm{str}}
=\operatorname{FeatureEncoder}\!\left(
\left[
\mathbf{x}_p^{\mathrm{cat}}
\Vert\mathbf{x}_p^{\mathrm{hour}}
\Vert\mathbf{x}_p^{\mathrm{week}}
\Vert\mathbf{x}_p^{\mathrm{geo}}
\right]
\right),
\end{equation}
where \(\mathbf{x}_p^{\mathrm{hour}}\) and
\(\mathbf{x}_p^{\mathrm{week}}\) denote the hour-bin and weekday
visitation-frequency distributions of POI \(p\), respectively,
while \(\mathbf{x}_p^{\mathrm{cat}}\) encodes its category and
\(\mathbf{x}_p^{\mathrm{geo}}\) denotes its geographic location
relative to the dataset center.

We concatenate the behavioral and structured representations as
\(\mathbf{y}_p=[\mathbf{h}_p^{\mathrm{beh}}/\sqrt{2}\Vert
\mathbf{h}_p^{\mathrm{str}}/\sqrt{2}]\) and encode the result through
the RQ-VAE encoder:
\begin{equation}
\mathbf{z}_p
=\operatorname{Encoder}(\mathbf{y}_p).
\end{equation}
Starting from \(\mathbf{r}_p^{(0)}=\mathbf{z}_p\), its \(L\)-level
residual quantizer selects the nearest codeword at each level and
updates the residual:
\begin{equation}
q_p^{(\ell)}
=\arg\min_{k}
\left\|
\mathbf{r}_p^{(\ell-1)}-\mathbf{e}_{k}^{(\ell)}
\right\|_2^2,
\qquad
\mathbf{r}_p^{(\ell)}
=\mathbf{r}_p^{(\ell-1)}
-\mathbf{e}_p^{(\ell)}.
\end{equation}
Here, \(\mathbf{e}_k^{(\ell)}\) denotes the \(k\)-th codeword in the
\(\ell\)-th codebook, and \(\mathbf{e}_p^{(\ell)}\) denotes the
codeword selected for POI \(p\) at level \(\ell\), indexed by
\(q_p^{(\ell)}\).
The resulting SID and quantized representation are
\begin{equation}
\operatorname{SID}(p)
=\left\langle
q_p^{(1)},\ldots,q_p^{(L)}
\right\rangle,
\qquad
\widehat{\mathbf{z}}_p
=\sum_{\ell=1}^{L}
\mathbf{e}_p^{(\ell)}.
\end{equation}
Successive codebooks progressively reduce the remaining quantization
error, while shared SID components reflect similarities among POIs.
For example, consider two cafes with similar morning visitation patterns
and SIDs \(\langle 12,7,31\rangle\) and \(\langle 12,7,18\rangle\).
Their shared components reflect similarities in venue category and
visitation time, while their final components distinguish the two POIs.

The quantized representation is decoded to reconstruct \(\mathbf{y}_p\):
\begin{equation}
\widetilde{\mathbf{y}}_p
=\operatorname{Decoder}\!\left(
\widehat{\mathbf{z}}_p
\right),
\end{equation}
where \(\operatorname{Decoder}(\cdot)\) denotes the RQ-VAE
decoder. Their discrepancy is measured by
\begin{equation}
\mathcal{L}_{\mathrm{rec}}
=\frac{1}{|\mathcal{P}|}\sum_{p\in\mathcal{P}}
\left\|
\mathbf{y}_p-\widetilde{\mathbf{y}}_p
\right\|_2^2.
\end{equation}
To align each residual with its selected codeword, we use the codebook
loss~\cite{oord2017vqvae,lee2022rqvae}
\begin{equation}
\mathcal{L}_{\mathrm{cb}}
=\frac{1}{|\mathcal{P}|L}
\sum_{p\in\mathcal{P}}\sum_{\ell=1}^{L}
\left\|
\operatorname{sg}\!\left[
\mathbf{r}_p^{(\ell-1)}
\right]-\mathbf{e}_p^{(\ell)}
\right\|_2^2,
\end{equation}
and the commitment loss
\begin{equation}
\mathcal{L}_{\mathrm{com}}
=\frac{1}{|\mathcal{P}|L}
\sum_{p\in\mathcal{P}}\sum_{\ell=1}^{L}
\left\|
\mathbf{r}_p^{(\ell-1)}
-\operatorname{sg}\!\left[
\mathbf{e}_p^{(\ell)}
\right]
\right\|_2^2,
\end{equation}
where \(\operatorname{sg}[\cdot]\) stops the gradient. The codebook loss
moves selected codewords toward their assigned residuals, while the
commitment loss keeps the encoder outputs close to the selected
codewords. The overall RQ-VAE objective is
\begin{equation}
\mathcal{L}_{\mathrm{RQ}}
=\lambda_{\mathrm{rec}}\mathcal{L}_{\mathrm{rec}}
+\lambda_{\mathrm{cb}}\mathcal{L}_{\mathrm{cb}}
+\lambda_{\mathrm{com}}\mathcal{L}_{\mathrm{com}}.
\end{equation}
Each codebook is initialized by applying K-means to its input residual,
beginning with \(\mathbf{z}_p\) at the first level~\cite{rajput2023tiger}.

For downstream POI prediction, we use the level-specific codebook vector
of each SID component instead of its discrete index. These vectors
preserve richer relations in the learned semantic space and, together
with straight-through estimation~\cite{oord2017vqvae}, allow downstream gradients to pass
through SID construction during end-to-end learning.
The resulting semantic POI representation is
\begin{equation}
\mathbf{h}_p^{\mathrm{sem}}
=\operatorname{SemanticEncoder}\!\left(
\left[
\mathbf{e}_p^{(1)}\Vert
\cdots\Vert
\mathbf{e}_p^{(L)}
\right]
\right),
\label{eq:semantic-poi-representation}
\end{equation}
which preserves the contribution of each residual level and provides
the semantic features used in subsequent POI prediction.

\subsection{Category-Guided POI Prediction}
\label{sec:category-guided-poi-prediction}

We combine the agentic category predictions in
Equation~\ref{eq:agent-category-prediction} with the semantic POI
representations in Equation~\ref{eq:semantic-poi-representation} for
fine-grained POI prediction. The predicted categories provide
category-level guidance, while the semantic representations encode
fine-grained relations among individual POIs.

The semantic POI representations of the check-ins before the target
are stacked chronologically as \(\mathbf{H}_{u,n}^{\mathrm{sem}}\) and
augmented with the positional embedding matrix
\(\mathbf{E}^{\mathrm{pos}}\). A causal
Transformer~\cite{vaswani2017attention,kang2018sasrec} encodes the
sequence, and its final valid state summarizes the trajectory:
\begin{equation}
\mathbf{h}_{u,n}^{\mathrm{traj}}
=\operatorname{Transformer}\!\left(
\mathbf{H}_{u,n}^{\mathrm{sem}}+\mathbf{E}^{\mathrm{pos}}
\right)_{\mathrm{last}}.
\end{equation}

We combine the trajectory state with the target time to form the
prediction query:
\begin{equation}
\mathbf{q}_{u,n}
=\operatorname{Feedforward}\!\left(
\mathbf{h}_{u,n}^{\mathrm{traj}},t_{u,n+1}
\right).
\end{equation}

For each POI \(p\in\mathcal{P}\), a correction encoder integrates the
user query, semantic POI representation, predicted Top-\(K\) categories,
multi-granularity graph information, and mobility statistics to produce
a score correction:
\begin{equation}
\delta_{u,n}(p)
=\operatorname{CorrectionEncoder}\!\left(
\mathbf{q}_{u,n},\mathbf{h}_{p}^{\mathrm{sem}},
\mathbf{g}_{u,n+1}^{K},\mathbf{G}_n,\mathbf{S}_{u,n}
\right).
\end{equation}
The POI score combines semantic relevance with this correction:
\begin{equation}
a_{u,n}(p)
=\frac{\mathbf{q}_{u,n}^{\top}
\mathbf{h}_{p}^{\mathrm{sem}}}{\sqrt{d_{\mathrm{sem}}}}
+\delta_{u,n}(p),
\end{equation}
where \(d_{\mathrm{sem}}\) is the dimension of the semantic
representation. We obtain the next POI probability as
\begin{equation}
P(p\mid u,n)
=\frac{\exp\!\left(a_{u,n}(p)\right)}
{\sum_{p'\in\mathcal{P}}
\exp\!\left(a_{u,n}(p')\right)}.
\end{equation}

Given the ground-truth next POI \(p_{u,n+1}\), the POI prediction loss is
\begin{equation}
\mathcal{L}_{\mathrm{POI}}
=-\sum_{(u,n)}\log P(p_{u,n+1}\mid u,n).
\end{equation}
The complete learning objective is
\begin{equation}
\mathcal{L}
=\mathcal{L}_{\mathrm{POI}}
+\mathcal{L}_{\mathrm{RQ}}
+\lambda_{\mathrm{i2v}}\mathcal{L}_{\mathrm{i2v}},
\end{equation}
where \(\mathcal{L}_{\mathrm{i2v}}\) denotes the Item2Vec
negative-sampling loss~\cite{barkan2016item2vec}.
We jointly optimize RQ-VAE, semantic POI representations, and the
category-guided POI predictor end-to-end. Gradients pass through quantization via
straight-through estimation. The predictor uses the agent's category
guidance and multi-granularity transition patterns to distinguish
individual POIs.

\section{Experiment}

\subsection{Experimental Setup}

\noindent{\textnormal{\bfseries Datasets.}}
We evaluate SHARP on Foursquare-NYC, Foursquare-TKY~\cite{yang2015foursquare},
and Gowalla-CA~\cite{cho2011gowalla}, three real-world check-in datasets.
Following prior studies~\cite{yang2022getnext,huang2024mtnet,zhong2025comapoi},
we remove users and POIs with a low number of check-ins and split each user's trajectories
chronologically, reserving the last trajectory for testing, the
penultimate for validation, and the rest for training.
Table~\ref{tab:dataset-statistics} reports the dataset statistics.
Each sample contains 30 historical and 29 context check-ins, followed
by the target check-in.

\begin{table}[t]
  \centering
  \caption{Statistics of the datasets used in our experiments.}
  \label{tab:dataset-statistics}
  \small
  \setlength{\tabcolsep}{4pt}
  \resizebox{\columnwidth}{!}{%
  \begin{tabular}{@{}ccccccc@{}}
    \toprule
    \textbf{Dataset} & \textbf{Users} & \textbf{Check-ins}
    & \textbf{POIs} & \textbf{Categories}
    & \begin{tabular}[c]{@{}c@{}}\textbf{Avg. POIs}\\\textbf{per Category}\end{tabular}
    & \begin{tabular}[c]{@{}c@{}}\textbf{Avg. Trajectory}\\\textbf{Length}\end{tabular} \\
    \midrule
    NYC & 988 & 135,666 & 5,092 & 207 & 24.60 & 137.31 \\
    TKY & 2,206 & 407,992 & 7,852 & 189 & 41.54 & 184.95 \\
    CA & 1,818 & 234,323 & 13,627 & 301 & 45.27 & 128.89 \\
    \bottomrule
  \end{tabular}%
  }
\end{table}

\noindent{\textnormal{\bfseries Baselines.}}
We compare SHARP with twelve baselines in three groups.
\textbf{(1) General sequential recommendation methods.}
SASRec~\cite{kang2018sasrec}, BERT4Rec~\cite{sun2019bert4rec},
DuoRec~\cite{qiu2022duorec}, TiCoSeRec~\cite{dang2023ticoserec},
MAERec~\cite{ye2023maerec}, and DiffuRec~\cite{li2023diffurec}.
\textbf{(2) Spatiotemporal POI prediction methods.}
GETNext~\cite{yang2022getnext}, POIGDE~\cite{yang2024poigde},
CrossDR~\cite{tao2023crossdr}, and MTNet~\cite{huang2024mtnet}.
\textbf{(3) LLM-based POI prediction methods.} LLM4POI~\cite{li2024llm4poi} and
CoMaPOI~\cite{zhong2025comapoi}.
More details are shown in Appendix~\ref{app:baselines}.

\noindent{\textnormal{\bfseries Evaluation Metrics.}}
We report Acc@5/10, NDCG@5/10~\cite{jarvelin2002ndcg}, and
MRR~\cite{voorhees1999mrr}. MRR averages the target's reciprocal rank
within the Top-10, assigning zero to targets outside this range.
Detailed metric definitions are provided in Appendix~\ref{app:evaluation-metrics}.

% Queue the wide results table before its discussion in the next subsection.
\begin{table*}[!t]
  \centering
  \caption{Performance comparison in Acc@k, NDCG@k, and MRR on three
  datasets. Bold and underlined values denote the best and second-best
  performance, respectively. Improvement is relative to the best baseline
  for each metric.}
  \label{tab:main-results}
  \small
  \setlength{\tabcolsep}{3.2pt}
  \resizebox{\textwidth}{!}{%
  \begin{tabular}{@{}lccccc ccccc ccccc@{}}
    \toprule
    \multirow{2}{*}{\textbf{Method}}
    & \multicolumn{5}{c}{\textbf{NYC (\%)}}
    & \multicolumn{5}{c}{\textbf{TKY (\%)}}
    & \multicolumn{5}{c}{\textbf{CA (\%)}} \\
    \cmidrule(lr){2-6}\cmidrule(lr){7-11}\cmidrule(lr){12-16}
    & ACC@5 & ACC@10 & NDCG@5 & NDCG@10 & MRR
    & ACC@5 & ACC@10 & NDCG@5 & NDCG@10 & MRR
    & ACC@5 & ACC@10 & NDCG@5 & NDCG@10 & MRR \\
    \midrule
    SASRec
    & 39.27 & 47.27 & 27.13 & 29.71 & 24.19
    & 32.77 & 41.48 & 23.97 & 26.82 & 22.25
    & 19.36 & 25.63 & 14.91 & 16.99 & 14.31 \\
    BERT4Rec
    & 38.26 & 46.15 & 27.08 & 29.65 & 24.45
    & 33.95 & 41.84 & 24.93 & 27.52 & 23.03
    & 21.29 & 27.39 & 15.99 & 17.97 & 15.04 \\
    DuoRec
    & 34.01 & 40.28 & 23.25 & 25.29 & 20.55
    & 23.07 & 29.78 & 17.55 & 19.75 & 16.63
    & 17.55 & 23.10 & 13.32 & 15.15 & 12.69 \\
    TiCoSeRec
    & 39.17 & 47.06 & 27.56 & 30.15 & 24.81
    & 27.74 & 35.49 & 20.56 & 23.08 & 19.22
    & 20.68 & 27.28 & 15.75 & 17.90 & 15.01 \\
    MAERec
    & 36.64 & 43.62 & 26.00 & 28.28 & 23.43
    & 30.69 & 38.94 & 22.28 & 24.94 & 20.60
    & 22.06 & 28.55 & 16.44 & 18.59 & 15.50 \\
    DiffuRec
    & 35.83 & 40.38 & 27.26 & 28.80 & 25.06
    & 32.50 & 38.17 & 25.12 & 26.99 & 23.44
    & 19.91 & 23.49 & 16.05 & 17.23 & 15.24 \\
    \midrule
    GETNext
    & 42.61 & 50.61 & 30.79 & 33.42 & 27.97
    & 38.53 & 46.42 & 28.62 & 31.20 & 26.41
    & 22.22 & 28.16 & 16.52 & 18.46 & 15.44 \\
    POIGDE
    & 33.40 & 37.85 & 25.14 & 26.61 & 22.99
    & 26.38 & 30.46 & 20.88 & 22.24 & 19.61
    & 20.52 & 25.69 & 16.72 & 18.42 & 16.14 \\
    CrossDR
    & 39.68 & 46.36 & 28.67 & 30.84 & 25.92
    & 30.51 & 38.80 & 22.02 & 24.72 & 20.35
    & 19.97 & 27.56 & 14.82 & 17.30 & 14.16 \\
    MTNet
    & 46.26 & 52.53 & 33.65 & 35.72 & 30.33
    & 41.43 & 49.09 & 30.74 & 33.26 & 28.26
    & 27.50 & 34.76 & 20.61 & 22.99 & 19.32 \\
    \midrule
    LLM4POI
    & 43.12 & 46.26 & 35.03 & 36.12 & 32.76
    & 32.73 & 37.76 & 26.30 & 28.04 & 24.96
    & 21.34 & 24.86 & 17.84 & 19.04 & 17.20 \\
    CoMaPOI
    & \underline{50.10} & \underline{58.10} & \underline{37.61}
    & \underline{40.22} & \underline{34.51}
    & \underline{43.11} & \underline{52.09} & \underline{31.88}
    & \underline{34.77} & \underline{29.33}
    & \underline{30.03} & \underline{37.84} & \underline{22.77}
    & \underline{25.29} & \underline{21.38} \\
    \midrule
    SHARP(ours)
    & \textbf{53.14} & \textbf{59.41} & \textbf{41.15}
    & \textbf{43.22} & \textbf{38.02}
    & \textbf{49.50} & \textbf{59.29} & \textbf{37.50}
    & \textbf{40.68} & \textbf{34.83}
    & \textbf{36.47} & \textbf{42.90} & \textbf{27.80}
    & \textbf{29.90} & \textbf{25.79} \\
    \rowcolor{gray!15}
    Improvement
    & 6.07\% & 2.25\% & 9.41\% & 7.46\% & 10.17\%
    & 14.82\% & 13.82\% & 17.63\% & 17.00\% & 18.75\%
    & 21.45\% & 13.37\% & 22.09\% & 18.23\% & 20.63\% \\
    \bottomrule
  \end{tabular}%
  }
\end{table*}

\noindent{\textnormal{\bfseries Implementation Details.}}
We use Llama3.1-8B~\cite{grattafiori2024llama3} as the backbone for both
agents, served through vLLM~\cite{kwon2023vllm}.
We predict Top-10 categories and retrieve graph information
within three hops. Item2Vec and RQ-VAE are pretrained only on the
training set before end-to-end optimization. We use
AdamW~\cite{loshchilov2019adamw} with batch size 8 and learning rates
of \(10^{-3}\) for the POI predictor and \(10^{-4}\) for RQ-VAE.
Training runs for at most 100 epochs, with early stopping after five
epochs without improvement in the mean validation NDCG@5 and MRR.
Test results use the checkpoint with the best validation score.
More implementation details are shown in Appendix~\ref{app:implementation-details}.

\subsection{Performance Comparison with Baselines}

\noindent{\textnormal{\bfseries Overall Performance.}}
We compare SHARP with twelve baselines on NYC, TKY, and CA in
Table~\ref{tab:main-results}. SHARP achieves the best results across
all datasets and metrics. Compared with the strongest baseline,
CoMaPOI, its relative Acc@5 improvements are 6.07\%, 14.82\%, and
21.45\%, respectively, while MRR improves by 10.17\%, 18.75\%, and
20.63\%. The consistent gains in Acc@k, NDCG@k, and MRR show that
SHARP both identifies more target POIs and places them higher in the
prediction list.

\noindent{\textnormal{\bfseries General Sequential Recommendation Methods.}}
The highest Acc@5 among general sequential methods is 39.27\% on
NYC, 33.95\% on TKY, and 22.06\% on CA, compared with SHARP's
53.14\%, 49.50\%, and 36.47\%, respectively. These gaps suggest that
learning item sequences without explicitly accounting for mobility
semantics limits the modeling of POI choices. SHARP connects these
choices to category-level intentions and time-conditioned transitions.

\noindent{\textnormal{\bfseries Spatiotemporal POI Prediction Methods.}}
MTNet outperforms all general sequential methods across every dataset
and metric, while GETNext consistently exceeds SASRec, supporting the
value of explicitly modeling spatiotemporal mobility patterns.
Nevertheless, SHARP improves Acc@5 over MTNet by 6.88, 8.07, and
8.97 percentage points on NYC, TKY, and CA, respectively. This
advantage suggests that category-level semantic reasoning provides
complementary information to learned spatiotemporal relations.

\noindent{\textnormal{\bfseries LLM-Based POI Prediction Methods.}}
CoMaPOI is the strongest baseline on every metric, indicating the
value of combining LLM semantic knowledge with collaborative
trajectory reasoning. On TKY, however, LLM4POI achieves an Acc@5 of
32.73\%, below MTNet's 41.43\%, suggesting that semantic reasoning
alone does not adequately model specific POI choices. SHARP improves
over CoMaPOI with
relative NDCG@5 gains of 9.41\%, 17.63\%, and 22.09\% across the
three datasets. These results support separating category reasoning
from specific POI prediction and connecting them through semantic
POI representations.

\begin{table}[!t]
  \centering
  \caption{Ablation results in Acc@10 and NDCG@10 on three datasets.
  Bold values denote the best performance.}
  \label{tab:ablation-results}
  \small
  \setlength{\tabcolsep}{5pt}
  \resizebox{\columnwidth}{!}{%
  \begin{tabular}{@{}lcc cc cc@{}}
    \toprule
    \textbf{Method}
    & \multicolumn{2}{c}{\textbf{NYC (\%)}}
    & \multicolumn{2}{c}{\textbf{TKY (\%)}}
    & \multicolumn{2}{c}{\textbf{CA (\%)}} \\
    \cmidrule(lr){2-3}\cmidrule(lr){4-5}\cmidrule(lr){6-7}
    & ACC@10 & NDCG@10 & ACC@10 & NDCG@10 & ACC@10 & NDCG@10 \\
    \midrule
    SHARP(ours)
    & \textbf{59.41} & \textbf{43.22}
    & \textbf{59.29} & \textbf{40.68}
    & \textbf{42.90} & \textbf{29.90} \\
    \midrule
    w/o DTM
    & 55.36 & 42.84 & 54.44 & 37.90 & 36.58 & 27.09 \\
    w/o MTCTG
    & 55.67 & 41.83 & 55.21 & 39.03 & 40.04 & 28.78 \\
    w/o SPR
    & 57.19 & 41.50 & 56.89 & 39.17 & 40.98 & 28.55 \\
    Only Agents
    & 52.13 & 38.65 & 50.86 & 35.76 & 38.45 & 27.19 \\
    Only POI Predictor
    & 46.15 & 34.39 & 53.49 & 37.87 & 29.65 & 22.07 \\
    \bottomrule
  \end{tabular}%
  }
\end{table}

\subsection{Ablation Study}

We evaluate five variants of SHARP to confirm our design. We remove dynamic trajectory
memory (\emph{w/o DTM}) and multi-granularity time-aware category
transition graph information (\emph{w/o MTCTG}) in two variants.
We replace semantic POI representations with ordinary trainable POI
embeddings (\emph{w/o SPR}).
We retain the trajectory memory agent and adapt the category predict
agent to directly predict POIs (\emph{Only Agents}). We also use only
semantic POI representations and the POI predictor without
guidance from the agent's category predictions (\emph{Only POI Predictor}).

As shown in Table~\ref{tab:ablation-results}, all variants underperform
SHARP on both metrics across all three datasets. Among the first three
variants, removing dynamic trajectory memory causes the largest Acc@10
reductions of
4.05, 4.85, and 6.32 percentage points on NYC, TKY, and CA, respectively,
highlighting the benefit of retaining rich historical information while
updating mobility patterns. Removing multi-granularity time-aware
category transition graph information also reduces both metrics.
Replacing semantic POI representations with ordinary embeddings
consistently lowers performance, with NYC NDCG@10 decreasing from
43.22\% to 41.50\%, supporting the value of shared semantic relations.
The single-branch variants further highlight the benefit of hierarchical
prediction. On TKY, Acc@10 falls
from 59.29\% to 50.86\% when the agents directly predict POIs and to
53.49\% when the POI predictor lacks agent category guidance, showing
that neither branch alone matches their combination.
% Together, these results support hierarchical prediction informed by
% dynamic trajectory memory and time-aware graphs.

\begin{figure}[t]
  \centering
  \includegraphics[width=\columnwidth]{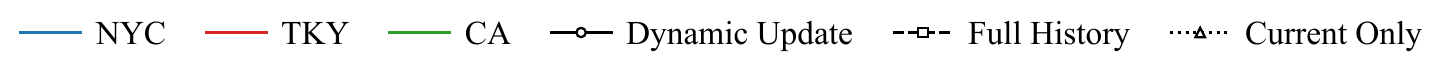}\par
  \smallskip
  \begin{minipage}[t]{0.49\columnwidth}
    \centering
    \includegraphics[width=\linewidth]{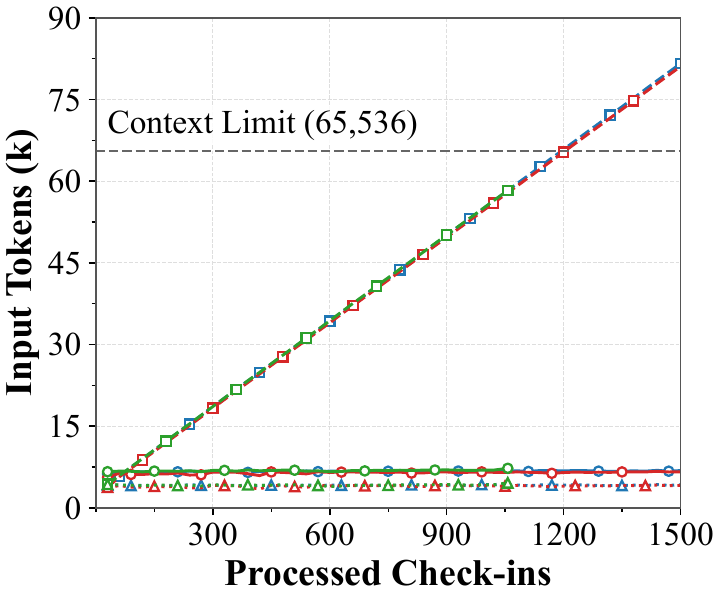}\par
    {\small\bfseries (a) Input Token Usage\par}
  \end{minipage}\hfill
  \begin{minipage}[t]{0.49\columnwidth}
    \centering
    \includegraphics[width=\linewidth]{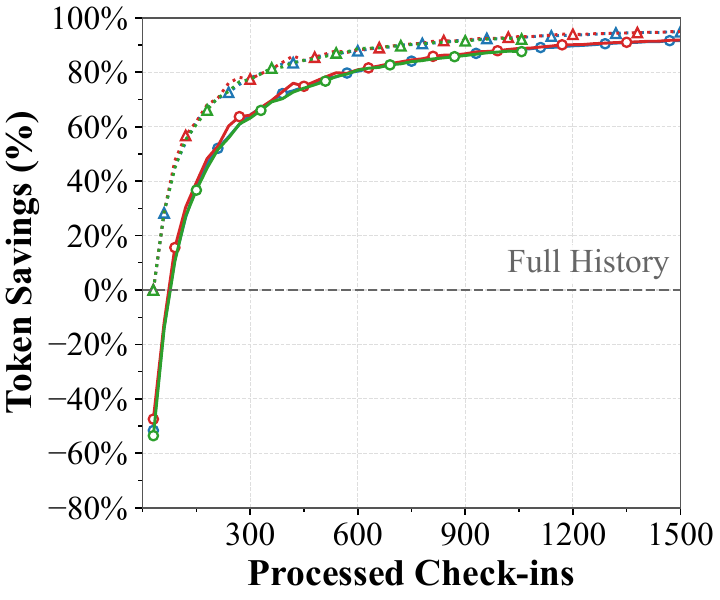}\par
    {\small\bfseries (b) Token Savings\par}
  \end{minipage}
  \caption{Input token usage and savings for updating the profile
  \(\mathbf{P}_{u,n}\).}
  \Description{Two plots with a shared legend compare input tokens per
  profile update and token savings as the number of processed check-ins
  increases for dynamic updating, full history, and current history only.}
  \label{fig:trajectory-memory-tokens}
\end{figure}

\subsection{Dynamic Trajectory Memory Analysis}

We examine the prediction benefits and token cost of dynamic trajectory
memory using three settings. \textbf{Dynamic Update} combines the previous
profile with the current historical trajectory and its statistics to
update \(\mathbf{P}_{u,n}\). \textbf{Current Only} generates a profile from only
the current historical trajectory and its statistics, while \textbf{Full
History} supplies all preceding historical check-ins at each step.
For token analysis, we select the user with the most check-ins in each
dataset and measure input tokens per profile update, with savings
relative to Full History.

\noindent{\textnormal{\bfseries Token Cost Saving.}}
As shown in Figure~\ref{fig:trajectory-memory-tokens}, Full History
requires an approximately linear increase in input tokens as check-ins
accumulate, whereas Dynamic Update remains stable at around 6--7k
tokens. As reported in Table~\ref{tab:trajectory-memory-tokens},
at 1500 check-ins, Full History exceeds
the configured context limit of 65,536 tokens on NYC and TKY, while
Dynamic Update saves 91.64\% and 91.86\%, respectively. On CA, the
saving reaches 86.14\% at 900 check-ins. Dynamic Update uses fewer
than 7k tokens in all three cases, with greater savings as check-ins
accumulate.

\begin{table}[t]
  \centering
  \caption{Input token counts, savings, and category prediction accuracy.}
  \label{tab:trajectory-memory-tokens}
  \small
  \setlength{\tabcolsep}{3pt}
  \resizebox{\columnwidth}{!}{%
  \begin{tabular}{@{}llrrrrccc@{}}
    \toprule
    \multirow{2}{*}{\textbf{Dataset}} & \multirow{2}{*}{\textbf{Method}}
    & \multicolumn{4}{c}{\textbf{History Check-ins}}
    & \multirow{2}{*}{\shortstack{\textbf{Token Saving}\\\textbf{(\%)}}}
    & \multicolumn{2}{c}{\textbf{Accuracy (\%)}} \\
    \cmidrule(lr){3-6}\cmidrule(lr){8-9}
    & & 30 & 300 & 900 & 1500 & & ACC@5 & ACC@10 \\
    \midrule
    \multirow{3}{*}{NYC} & Dynamic Update & 6,227 & 6,623 & 6,738 & 6,822 & 91.64 & \textbf{60.63} & \textbf{67.41} \\
        & Current Only & 4,107 & 4,108 & 4,163 & 4,212 & 94.84 & 48.08 & 58.10 \\
        & Full History & 4,107 & 18,552 & 50,015 & 81,588 & 0 & -- & -- \\
    \midrule
    \multirow{3}{*}{TKY} & Dynamic Update & 5,537 & 6,543 & 6,570 & 6,596 & 91.86 & \textbf{73.39} & \textbf{79.51} \\
        & Current Only & 3,755 & 4,107 & 4,079 & 4,093 & 94.95 & 63.28 & 71.67 \\
        & Full History & 3,755 & 18,278 & 49,676 & 81,038 & 0 & -- & -- \\
    \midrule
    \multirow{3}{*}{CA} & Dynamic Update & 6,616 & 6,847 & 6,943 & -- & 86.14 & \textbf{44.77} & \textbf{54.46} \\
       & Current Only & 4,309 & 4,199 & 4,242 & -- & 91.53 & 32.07 & 40.76 \\
       & Full History & 4,309 & 18,575 & 50,104 & -- & 0 & -- & -- \\
    \bottomrule
  \end{tabular}%
  }
\end{table}

\begin{figure}[t]
  \centering
  \begin{minipage}[t]{0.49\columnwidth}
    \centering
    \includegraphics[width=\linewidth]{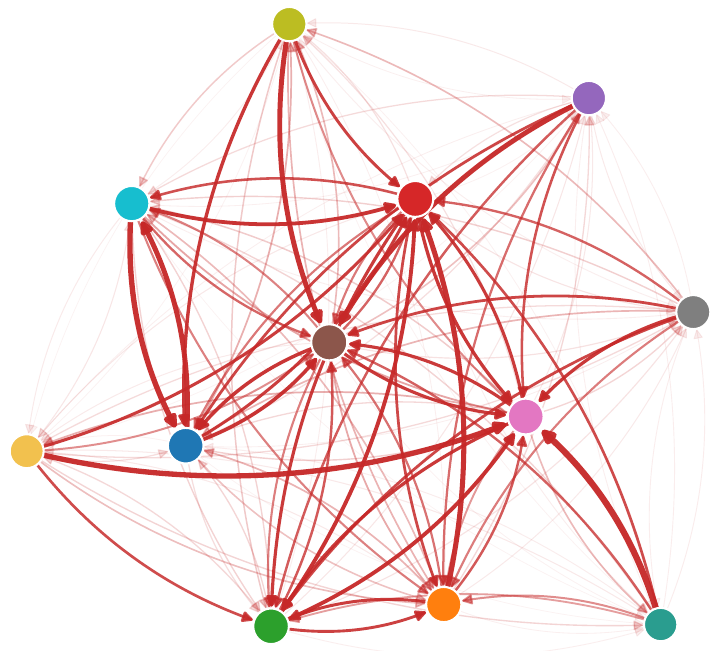}\par
    {\small\bfseries (a) Global\par}
  \end{minipage}\hfill
  \begin{minipage}[t]{0.49\columnwidth}
    \centering
    \includegraphics[width=\linewidth]{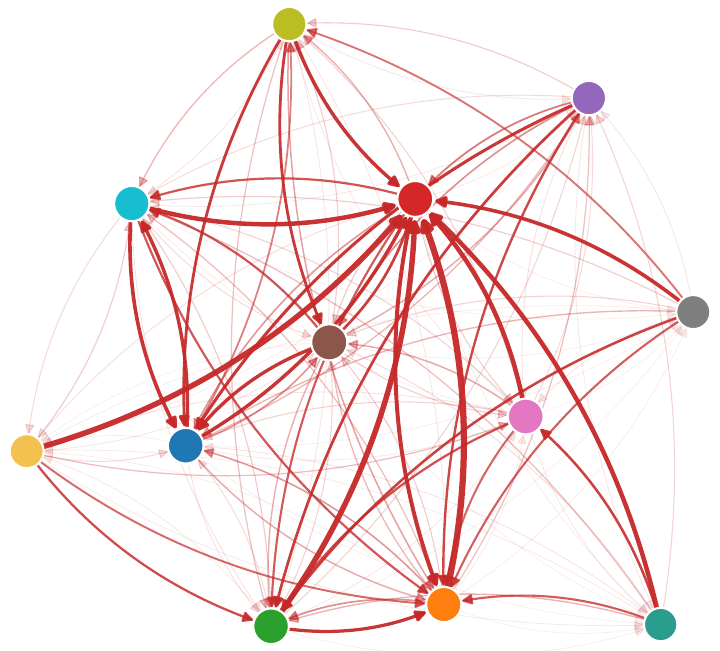}\par
    {\small\bfseries (b) Weekday Morning\par}
  \end{minipage}\par
  \smallskip
  \begin{minipage}[t]{0.49\columnwidth}
    \centering
    \includegraphics[width=\linewidth]{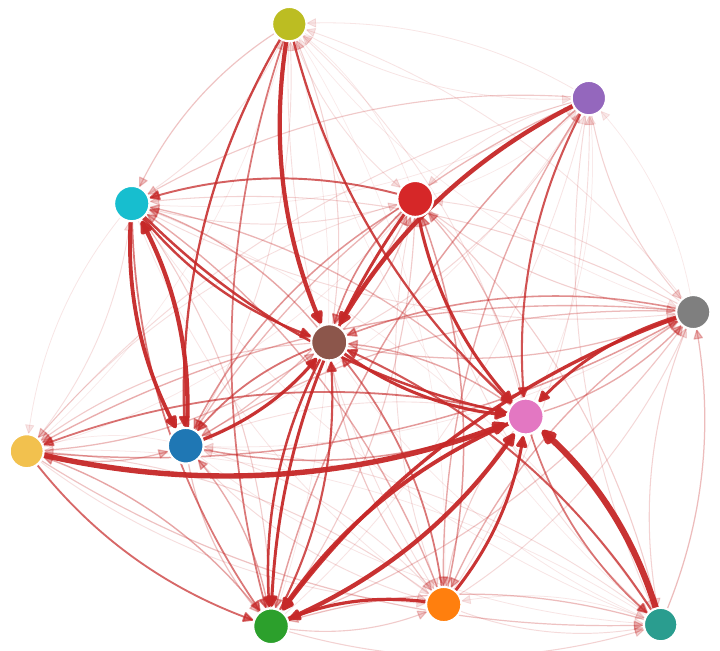}\par
    {\small\bfseries (c) Weekday Evening\par}
  \end{minipage}\hfill
  \begin{minipage}[t]{0.49\columnwidth}
    \centering
    \includegraphics[width=\linewidth]{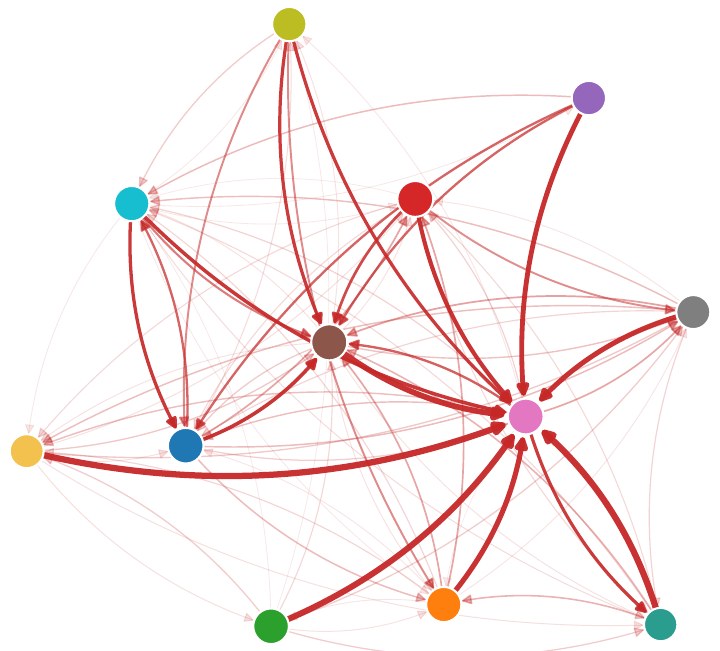}\par
    {\small\bfseries (d) Weekend Night\par}
  \end{minipage}\par
  \smallskip
  \includegraphics[width=\columnwidth]{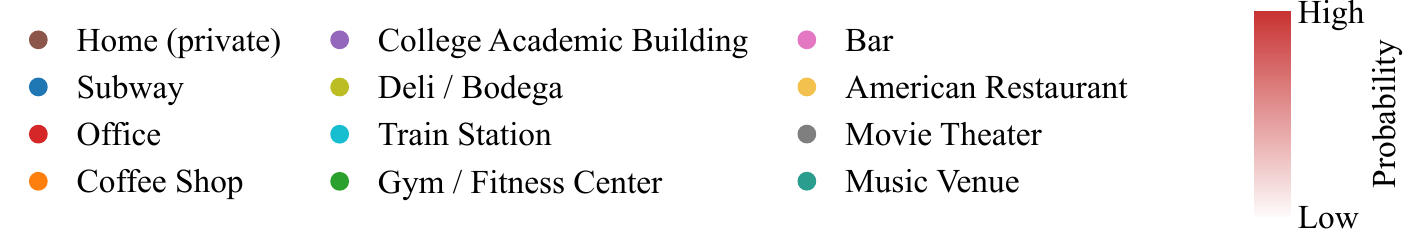}
  \caption{Category transition graphs under different time conditions on NYC.}
  \Description{Four directed category graphs with shared node positions and
  colors show global, weekday morning, weekday evening, and weekend night
  transitions. A shared legend below identifies the categories and relative
  transition probabilities within each graph.}
  \label{fig:temporal-category-graphs}
\end{figure}

\noindent{\textnormal{\bfseries Accuracy Improvements.}}
Compared with Current Only, Dynamic Update uses about 2.5--2.7k
additional tokens for the previous mobility profile and cumulative
history statistics, while both settings maintain stable input lengths.
As reported in Table~\ref{tab:trajectory-memory-tokens},
SHARP improves Acc@5 over Current Only by 12.55, 10.11, and 12.70
percentage points on NYC, TKY, and CA, respectively, with consistent
gains in Acc@10. Retaining and updating earlier mobility patterns thus
improves category prediction with few extra tokens.

\subsection{Multi-Granularity Category Transition Graph Analysis}

We examine the time-conditioned category transition patterns illustrated
in Figure~\ref{fig:intro-comparison}(c).
Using twelve selected categories from NYC, we compare the global
graph with three conditional graphs for weekday morning, weekday
evening, and weekend night.

Figure~\ref{fig:temporal-category-graphs} shows overlapping transition
patterns in the global graph, whereas the conditional views reveal
more distinct temporal structure. Transitions toward Office are
prominent on weekday mornings, while transitions toward Bar become
more prominent in the evening and on weekend nights.
As shown in Table~\ref{tab:temporal-category-transition-probabilities},
Coffee Shop \(\rightarrow\) Office has a global
probability of 31.29\%, compared with 58.12\% on weekday mornings and
9.20\% on weekday evenings. In contrast, Coffee Shop \(\rightarrow\)
Bar rises from 0.92\% on weekday mornings to 25.29\% on weekday
evenings and 53.12\% on weekend nights, despite a global probability
of only 12.94\%. Aggregating transitions across temporal conditions
therefore smooths out time-specific variations, obscuring dominant
category transition patterns within particular time periods.
The conditional graphs better capture these temporal patterns and
provide more informative category transitions for category prediction.

\begin{table}[t]
  \centering
  \caption{Selected category transition probabilities (\%) on NYC.
  Bold values indicate row maxima.}
  \label{tab:temporal-category-transition-probabilities}
  \small
  \setlength{\tabcolsep}{3pt}
  \resizebox{\columnwidth}{!}{%
  \begin{tabular}{@{}lcccc@{}}
    \toprule
    \textbf{Transition} & \textbf{Global}
    & \shortstack{\textbf{Weekday}\\\textbf{Morning}}
    & \shortstack{\textbf{Weekday}\\\textbf{Evening}}
    & \shortstack{\textbf{Weekend}\\\textbf{Night}} \\
    \midrule
    Coffee Shop \(\rightarrow\) Office
    & 31.29 & \textbf{58.12} & 9.20 & 12.50 \\
    Coffee Shop \(\rightarrow\) Bar
    & 12.94 & 0.92 & 25.29 & \textbf{53.12} \\
    Home \(\rightarrow\) Office
    & 12.16 & \textbf{21.35} & 6.27 & 3.90 \\
    Home \(\rightarrow\) Subway
    & 20.18 & \textbf{28.43} & 12.28 & 7.79 \\
    Home \(\rightarrow\) Bar
    & 17.83 & 4.79 & 25.56 & \textbf{59.74} \\
    Office \(\rightarrow\) Bar
    & 18.88 & 5.04 & 25.74 & \textbf{41.51} \\
    \bottomrule
  \end{tabular}%
  }
\end{table}

\subsection{Semantic Structure Analysis of SIDs}

\begin{figure}[t]
  \centering
  \edef\sidpanelheight{\the\dimexpr0.42\columnwidth\relax}
  \begin{minipage}[t]{0.55\columnwidth}
    \centering
    \includegraphics[height=\sidpanelheight]{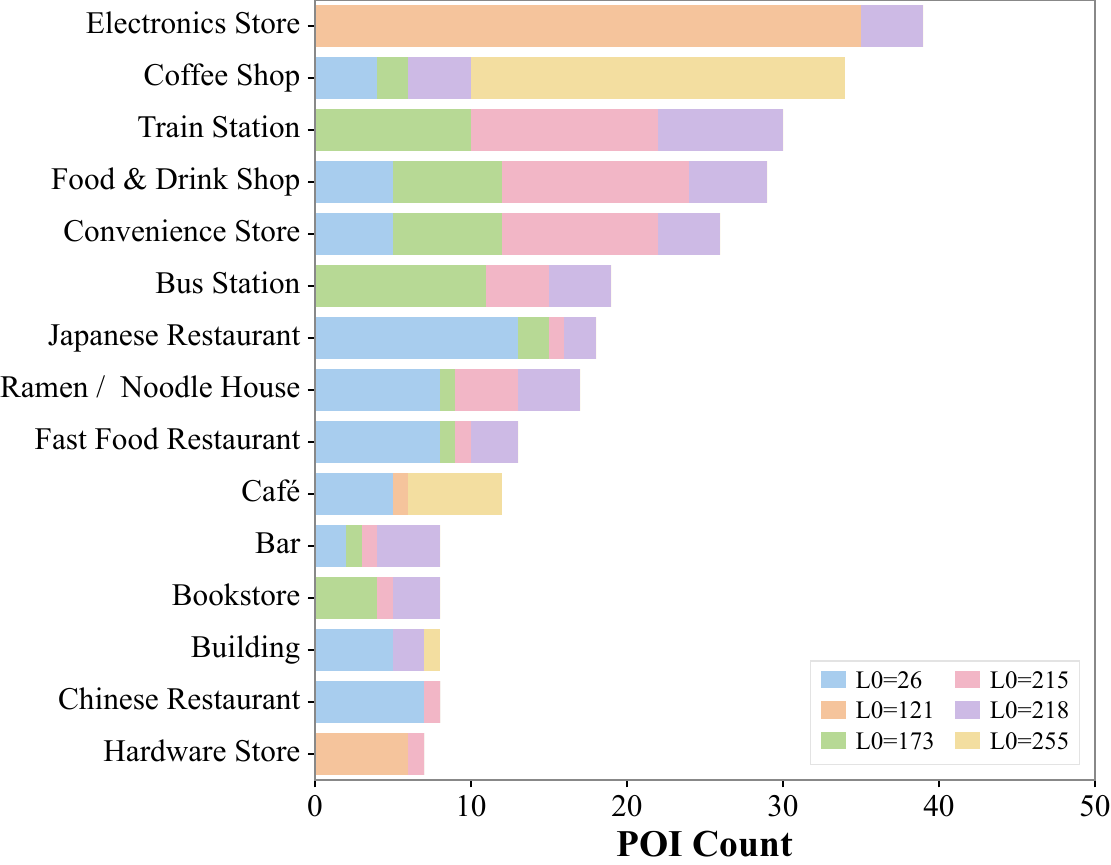}\par
    {\small\bfseries (a) Category Distribution\par}
  \end{minipage}\hfill
  \begin{minipage}[t]{0.44\columnwidth}
    \centering
    \includegraphics[height=\sidpanelheight]{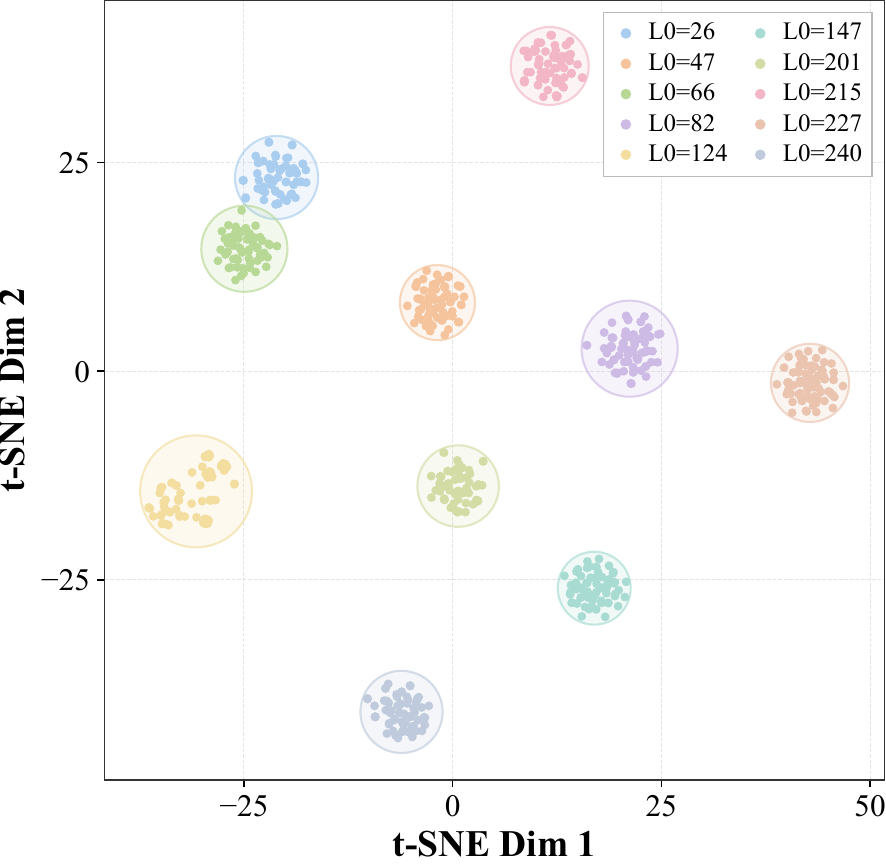}\par
    {\small\bfseries (b) t-SNE Visualization\par}
  \end{minipage}
  \caption{Semantic structure of SIDs on TKY.}
  \Description{The left panel shows category counts across six selected
  first-level SID codes. The right panel shows a t-SNE projection of
  concatenated codebook vectors for ten selected first-level codes,
  with translucent circles outlining each group. Shared codes use the
  same colors in both panels.}
  \label{fig:sid-semantic-structure}
\end{figure}

We analyze the category associations and codebook-space structure of
SIDs on TKY. Figure~\ref{fig:sid-semantic-structure}(a) shows category
counts within six selected first-level SID groups. Several prefixes
contain functionally related categories. For example, L0=26 groups
dining-related categories such as Japanese Restaurant and Ramen / Noodle
House, L0=255 groups Coffee Shop and Caf\'{e}, and L0=121 groups
Electronics Store and Hardware Store.
These associations indicate that shared SID prefixes
retain category-related semantics. Meanwhile, categories such as
Train Station occur across multiple prefixes. This distribution is
consistent with SID construction jointly encoding category, temporal,
geographic, and sequential information. POIs in the same category can
differ in these other characteristics, while POIs from different
categories can share functional or behavioral similarities. The
many-to-many relationship between categories and SID prefixes thus
allows shared category semantics and within-category differences to
be represented together. Combined with the agent's category guidance
and trajectory context, these distinctions help the POI predictor
differentiate mobility patterns among POIs in the same category.

Figure~\ref{fig:sid-semantic-structure}(b) presents a
t-SNE~\cite{maaten2008tsne} projection
of the concatenated codebook vectors that the Semantic Encoder maps
into semantic POI representations. POIs sharing the same first-level code
form compact groups, with clear separation between different groups
and variation among POIs within each group. This organization
reflects the coarse grouping induced by shared codewords while
retaining distinctions among individual POIs. Together, the two
views show that SIDs combine category-related semantic associations
with a structured representation space, providing distinctions beyond
category labels for fine-grained POI prediction.

\section{Conclusion}

We presented SHARP, an agentic hierarchical framework that connects
category-level reasoning with semantic POI representations for next
POI prediction. The representations and category-guided POI predictor
are jointly optimized end-to-end. Dynamic trajectory memory preserves
earlier mobility patterns and tracks behavioral changes, while
multi-granularity time-aware category transition graphs provide global
and time-conditioned patterns for both prediction stages. Experiments
on three real-world datasets demonstrate SHARP's predictive advantages
and substantial token savings through dynamic trajectory memory.

\bibliographystyle{ACM-Reference-Format}
\bibliography{references}
\appendix

\section*{Appendix}

\section{Experimental Details}
\label{app:experimental-details}

\subsection{Baseline Descriptions}
\label{app:baselines}

\noindent{\textnormal{\bfseries General Sequential Recommendation Methods.}}
SASRec~\cite{kang2018sasrec} uses causal self-attention to model
dependencies in a user's interaction sequence. It selects relevant
historical interactions at each position to predict the next item,
capturing both recent and longer-term preferences.

BERT4Rec~\cite{sun2019bert4rec} learns sequence representations with
a bidirectional Transformer. It predicts masked items from their
surrounding context during training and predicts the next item by
appending a masked position to the observed sequence.

DuoRec~\cite{qiu2022duorec} addresses representation degeneration
through contrastive regularization. It creates sequence views using
dropout and treats sequences with the same target item as positive
pairs, encouraging semantically consistent and well-distributed
representations.

TiCoSeRec~\cite{dang2023ticoserec} incorporates time intervals into
sequence augmentation. It applies interval-aware cropping, reordering,
masking, substitution, and insertion to produce more temporally uniform
sequences, reducing the effects of preference drift in contrastive
sequence modeling.

MAERec~\cite{ye2023maerec} augments sequential prediction with a graph
masked autoencoder. It adaptively masks and reconstructs item relations
to extract global transition information, providing self-supervision
for sequence representations under sparse and noisy interactions.

DiffuRec~\cite{li2023diffurec} formulates next-item prediction as
conditional denoising of item representations. It adds noise to target
embeddings during training and reconstructs a target representation
from Gaussian noise conditioned on the interaction history during
inference, modeling diverse user interests.

\noindent{\textnormal{\bfseries Spatiotemporal POI Prediction Methods.}}
GETNext~\cite{yang2022getnext} constructs a global trajectory flow map
to capture collective POI transitions. Its graph-enhanced Transformer
combines these patterns with user preferences, spatiotemporal context,
and time-aware category embeddings for next POI prediction.

POIGDE~\cite{yang2024poigde} models continuous changes in mobility
representations through graph differential equations. A time-serial
graph and interval-aware self-attention capture temporal evolution,
while Siamese learning aligns the predicted representation with the
target POI embedding.

CrossDR~\cite{tao2023crossdr} learns disentangled spatial and temporal
representations for next POI prediction. By separately modeling these
influences and combining their representations, it distinguishes
different factors underlying users' check-in transitions.

MTNet~\cite{huang2024mtnet} organizes check-ins into a Mobility Tree
with time-slot nodes at multiple granularities. Information propagates
from leaf nodes to the root to capture personalized temporal preferences,
and multitask training supports learning across temporal levels.

\noindent{\textnormal{\bfseries LLM-Based POI Prediction Methods.}}
LLM4POI~\cite{li2024llm4poi} adapts pretrained LLMs to next POI
prediction using textual representations of check-ins and contextual
information. This formulation preserves heterogeneous information and
uses the model's commonsense knowledge to interpret mobility context.

CoMaPOI~\cite{zhong2025comapoi} coordinates a Profiler, Forecaster,
and Predictor for next POI prediction. The Profiler converts mobility
information into language descriptions, the Forecaster refines the POI
candidate space, and the Predictor combines their outputs to predict
the next POI.

\subsection{Evaluation Metrics}
\label{app:evaluation-metrics}

We report Acc@5/10, NDCG@5/10, and MRR. Acc@\(k\) measures the
fraction of examples whose target POI appears in the Top-\(k\)
predictions. With one target per example, NDCG@\(k\) assigns a hit
at rank \(r\leq k\) the value \(1/\log_2(r+1)\), and zero
otherwise~\cite{jarvelin2002ndcg}. MRR assigns the value \(1/r\)
when the target appears within the Top-10, and zero
otherwise~\cite{voorhees1999mrr}. We average each metric over all
evaluation examples. Higher values indicate better performance.

\subsection{Implementation Details}
\label{app:implementation-details}

We use Foursquare-NYC, Foursquare-TKY~\cite{yang2015foursquare}, and
Gowalla-CA~\cite{cho2011gowalla}. Each sample contains 30 historical
and 29 context check-ins followed by the target check-in. For each
user, trajectories are ordered chronologically, with the last used for
testing, the penultimate for validation, and the remainder for training.

Experiments are conducted on two NVIDIA A100 80\,GB GPUs.
Both agents use Llama3.1-8B~\cite{grattafiori2024llama3}, served through
vLLM~\cite{kwon2023vllm}. The trajectory memory agent performs
long-term mobility profile updating and context interpretation; the
category predict agent produces Top-10 category predictions. Graph
retrieval uses a maximum hop depth of three.

Item2Vec and RQ-VAE are pretrained on the training set to stabilize
joint optimization. Each residual codebook is initialized by K-means
on its input residual. The semantic POI representations and the
category-guided POI predictor are then optimized jointly end to end
using AdamW~\cite{loshchilov2019adamw} with batch size 8 and weight
decay \(10^{-4}\). The learning rates are \(10^{-3}\) for the POI
predictor and \(10^{-4}\) for RQ-VAE. The maximum input sequence
length is 256. The Correction Encoder uses a hidden dimension of 64
and a dropout rate of 0.2. The loss weights are
\(\lambda_{\mathrm{rec}}=0.1\),
\(\lambda_{\mathrm{cb}}=1.0\),
\(\lambda_{\mathrm{com}}=0.25\), and
\(\lambda_{\mathrm{i2v}}=0.05\).
Training runs for at most 100 epochs with random seed 42.
The validation score is the mean of NDCG@5 and MRR;
training stops after five epochs without improvement. We evaluate
the checkpoint with the best validation score on the test set.

\section{Agent Prompts}
\label{app:agent-prompts}

The following templates specify the two stages of the trajectory memory
agent and the category predict agent. Braced placeholders are replaced
with the corresponding trajectory information at each step. All check-in
sequences are ordered chronologically. 

\lstdefinestyle{agentprompt}{
  basicstyle=\ttfamily\footnotesize,
  columns=fullflexible,
  keepspaces=true,
  breaklines=true,
  breakatwhitespace=false,
  showstringspaces=false,
  numbers=none,
  frame=none,
  aboveskip=0pt,
  belowskip=0pt
}
\newtcblisting{agentprompt}[1]{
  enhanced jigsaw,
  breakable,
  listing only,
  listing options={style=agentprompt},
  title={#1},
  fonttitle=\bfseries\small,
  coltitle=white,
  colbacktitle=gray!75!black,
  colback=gray!8,
  colframe=gray!65!black,
  boxrule=0.5pt,
  arc=2pt,
  left=4pt,right=4pt,top=4pt,bottom=4pt,
  before skip=\smallskipamount,
  after skip=\smallskipamount
}

\subsection{Long-Term Mobility Profile Updating}
\label{app:profile-prompt}

\begin{agentprompt}{Long-Term Mobility Profile Updating}
## System
You are the trajectory memory agent for next POI category prediction. Update the user's long-term mobility profile using the historical category-time sequence, statistical summaries, and previous profile.
## Instructions
1. Synthesize the ordered historical check-ins with both current and cumulative statistics. Use cumulative information for recurring patterns and current information for behavioral changes.
2. Use the previous profile as a reference. Preserve supported patterns and revise descriptions when new trajectories indicate changes. If no previous profile is supplied, initialize it from the available history.
3. Infer age and identity only as behavioral hypotheses, not demographic facts. Use cautious language when the supporting information is indirect.
4. Describe commuting, routines, and spatial activity using temporal and geographic statistics. Describe social habits only through observed dining, nightlife, and entertainment patterns.
5. Describe category transitions using recurring ordered transitions and their temporal settings. Do not use target or future check-ins.
6. Interpret hour bins as late_night [00:00,06:00), morning [06:00,11:00), noon [11:00,14:00), afternoon [14:00,18:00), evening [18:00,22:00), and night [22:00,24:00). Use day-of-week statistics for weekdays and week-type statistics for weekday/weekend comparisons.
## User
Update all seven profile dimensions from the following inputs. Summarize mobility patterns rather than repeating the statistics or copying the previous profile.
User: {user_id}
Trajectory: {user_sample_index}
Historical category-time sequence: {current_history_trajectory}
Current historical statistics: {current_history_statistics}
Cumulative historical statistics: {cumulative_history_statistics}
Previous long-term mobility profile: {previous_profile_or_null}
## Output
Return exactly the following seven lines. Each value must be one English sentence of at most 25 words. Do not include confidence scores, lists, explanations, or semicolons within a description.
age={age_description}
identity={identity_description}
commuting_pattern={commuting_description}
daily_routine={routine_description}
spatial_activity_range={spatial_description}
social_habits={social_description}
category_transition_pattern={transition_description}
\end{agentprompt}

\subsection{Context Interpretation}
\label{app:context-prompt}

\begin{agentprompt}{Context Interpretation}
## System
You are the trajectory memory agent for next POI category prediction. Interpret the user's recent context using the context category-time sequence, statistical summaries, and updated long-term mobility profile.
## Instructions
1. Use the long-term profile as a baseline for comparison, not as a fact to copy. Combine its descriptions with the supplied category preferences and transition patterns.
2. Analyze category trends, temporal behavior, spatial patterns, and ordered transitions together. Emphasize the most recent check-ins while using the complete context to distinguish recurring behavior from isolated events.
3. Consider the intervals between check-ins. Large time gaps may separate activity episodes; adjacent records do not necessarily describe one continuous trip.
4. Interpret hour bins as late_night [00:00,06:00), morning [06:00,11:00), noon [11:00,14:00), afternoon [14:00,18:00), evening [18:00,22:00), and night [22:00,24:00).
5. Compare recent behavior with the profile using both descriptions and statistical comparisons. If no profile is available, state that no historical baseline is available and do not invent changes relative to history.
6. Do not infer new demographic facts, personality, or actual sociability. Describe recent behavior without predicting the next category or POI.
## User
Analyze all five context dimensions from the following inputs. Describe recent mobility patterns and their relationship to the long-term profile rather than repeating raw statistics.
User: {user_id}
Trajectory: {user_sample_index}
Updated long-term mobility profile: {profile_descriptions}
Long-term category preferences: {cumulative_category_preferences}
Recurring historical transitions: {cumulative_transition_patterns}
Context category-time sequence and gaps: {ordered_context}
Context category, temporal, spatial, and transition statistics: {context_statistics}
Statistical comparison with the profile: {profile_comparison}
## Output
Return exactly the following five lines. Each value must be one concise English sentence. Do not include confidence scores, lists, explanations, or semicolons within a description.
category_trend={category_trend_description}
temporal_state={temporal_state_description}
spatial_state={spatial_state_description}
activity_intent={activity_intent_description}
profile_alignment={profile_alignment_description}
\end{agentprompt}

\subsection{Category Prediction}
\label{app:category-prompt}

\begin{agentprompt}{Category Prediction}
## System
You are the category predict agent for next POI prediction. Predict the Top-{K} next categories using the trajectory memory, context sequence, target time, and multi-granularity category transition information.
## Instructions
1. Select only from the supplied dataset category catalog. Never invent, rename, or merge categories. Categories without historical visits or graph support remain eligible; counts and ratios are supporting information, not selection constraints.
2. Prioritize recent context behavior and target time, then consider graph information, category statistics, and the long-term mobility profile. Treat inferred age and identity as weak supporting information rather than facts.
3. Use the ordered category sequence, check-in times, and time gaps together with the context descriptions to interpret recent transitions and temporal behavior.
4. The graph source is the final context category. The graph snapshot is bounded by the supplied observation boundary. All hops use the same snapshot and time condition.
5. Compare the global, week-type, hour-bin, and joint views. Specific time conditions are informative when supported by sufficient transitions; the global view provides broader transition patterns.
6. One-hop information contains transition counts and conditional probabilities. Multi-hop information contains destination categories, shortest positive hops, and normalized scores. Path probabilities are products of conditional probabilities, and paths reaching the same category at its shortest hop are combined. Longer paths do not add further contributions. Returns to the source category remain valid.
7. Interpret normalized multi-hop scores as structural support, not literal next-check-in probabilities. Consider them alongside one-hop transitions. An empty graph result means no observed support, not impossibility.
## User
Predict the Top-{K} categories from most likely to least likely using the following inputs.
Long-term mobility profile: {profile_descriptions}
Historical category and transition statistics: {profile_statistics}
Context state: {context_descriptions}
Context statistics: {context_statistics}
Ordered context check-ins: {ordered_context}
Final context category: {last_observed_context_category}
Target time and temporal attributes: {target_time}
Dataset category catalog and visit statistics: {eligible_categories}
Graph observation boundary and hop limit: {graph_observation_boundary}
One-hop transitions for the four graph views: {macro_graph_outgoing_transitions}
Multi-hop transitions for the four graph views: {macro_graph_multi_hop_transitions}
## Output
Return exactly {K} prediction lines, ordered from most likely to least likely. Each value must exactly match a category name in the supplied catalog. Do not include explanations or POI predictions.
prediction_1={category_name_1}
...
prediction_{K}={category_name_K}
\end{agentprompt}

\end{document}